\documentclass{sig-alternate-05-2015}
\usepackage{graphicx}
\usepackage{color}

 \usepackage{caption}
  \DeclareCaptionType{copyrightbox}
  \usepackage{subcaption}
  \usepackage{epstopdf}

\newenvironment{titemize} % Same as itemize, but with minimal vspace
        {\begin{list}{\labelitemi}{
		\setlength{\topsep}{0pt}
                \setlength{\parskip}{0pt}
                \setlength{\itemsep}{0pt}
                \setlength{\parsep}{0pt}
		\setlength{\leftmargin}{.5\parindent}
		\setlength{\labelwidth}{\parindent}
        }}
        {\end{list}}

\begin{document}

% Copyright

\CopyrightYear{2016} 
\setcopyright{acmlicensed}
\conferenceinfo{CIKM'16 ,}{October 24 - 28, 2016, Indianapolis, IN, USA}
\isbn{978-1-4503-4073-1/16/10}\acmPrice{\$15.00}
\doi{http://dx.doi.org/10.1145/2983323.2983692}

\clubpenalty=10000 
\widowpenalty = 10000

% DOI
\doi{http://dx.doi.org/10.1145/2983323.2983692}

% ISBN
\isbn{123-4567-24-567/08/06}

\title{Understanding Stability of Noisy Networks\\ through Centrality Measures and Local Connections}

\numberofauthors{4} 

\author{
\alignauthor
Vladimir Ufimtsev\\
\affaddr{Dept. of CS}\\
       \affaddr{Univ. of Nebraska at Omaha}\\
       \affaddr{NE 68182, USA}\\
        \email{vufimtsev@unomaha.edu}
% 2nd. author
\alignauthor
Soumya Sarkar\\
       \affaddr{Dept. of CSE}\\
\affaddr{IIT Kharagpur, India -- 721302}\\
       \email{soumya015@iitkgp.ac.in}
       \and
% 3rd. author
\alignauthor Animesh Mukherjee \\  
\affaddr{Dept. of CSE}\\
\affaddr{IIT Kharagpur, India -- 721302}\\
       \email{animeshm@cse.\\iitkgp.ernet.in}
  % use '\and' if you need 'another row' of author names
% 4th. author
\alignauthor Sanjukta Bhowmick\\
\affaddr{Dept. of CS}\\
   \affaddr{Univ. of Nebraska at Omaha}\\
   \affaddr{NE 68182, USA}\\
       \email{sbhowmick@unomaha.edu}
}
\maketitle
\begin{abstract}
 Networks created from real-world data contain some inaccuracies or noise, manifested as small changes in the network structure.  An important question is whether these small changes can significantly affect the analysis results.
 
 In this paper, we study the effect of noise in {\em changing ranks of the high centrality vertices.} We compare, using the Jaccard Index (JI), how many of the top-$k$ high centrality nodes from the original network are also part of the top-$k$ ranked nodes from the noisy network. We deem a network as stable if the JI value is high.
  
We observe two  features that affect the stability. First, {\em the stability is dependent on the number of top-ranked vertices considered}. When the vertices are ordered according to their centrality values, they group into clusters. 
Perturbations to the network can change the relative ranking within the cluster, but vertices rarely move from one cluster to another.
Second, {\em the stability is dependent on the local connections of the high ranking vertices}. The network is highly stable if the high ranking vertices are connected to each other.

Our findings show that the stability of a network is affected by the local properties of high centrality vertices, rather than the global properties of the entire network. 
Based on these local properties we can identify the stability of a network, without explicitly applying a noise model. 
\end{abstract}

\begin{CCSXML}
<ccs2012>
<concept>
<concept_id>10003033.10003083.10003095</concept_id>
<concept_desc>Networks~Network reliabihttps://preview.overleaf.com/public/jwnzzpmcybsk/images/5619b2e4151c9776c5077cee84255d223d0f1916.jpeglity</concept_desc>
<concept_significance>500</concept_significance>
</concept>
<concept>
<concept_id>10003752.10003809.10003635</concept_id>
<concept_desc>Theory of computation~Graph algorithms analysis</concept_desc>
<concept_significance>300</concept_significance>
</concept>
<concept>
<concept_id>10003752.10010061.10010069</concept_id>
<concept_desc>Theory of computation~Random network models</concept_desc>
<concept_significance>300</concept_significance>
</concept>
</ccs2012>
\end{CCSXML}

\ccsdesc[500]{Networks~Network reliability}
\ccsdesc[300]{Theory of computation~Graph algorithms analysis}
\ccsdesc[300]{Theory of computation~Random networks}

%
%  Use this command to print the description
%
\printccsdesc

\keywords{betweenness; closeness; stability; noise; rich-club}

\section{Introduction}
Network analysis is a very efficient tool for understanding complex systems of interacting entities that arise in diverse applications. Analysis of the network models provide insights to the properties of the underlying systems.

However, measurements of real-world systems is influenced by the experimental setup. Modeling of this data as networks is also affected by subjective choices. Given these uncertainties in data collection, any network created from real-world data will contain some inaccuracy (or noise). Noise in networks is manifested in the form of extra or missing edges.

{\bf Measuring the effect of noise:} In recent studies, researchers perturb the network by adding/deleting a specified percentage of edges (noise level). Then they observe by how much the properties of the network alter and correlate this change in the properties with the noise level. A network is deemed stable if the noise does not affect the properties.  Most of these studies focus on vertex-based measurements including centrality metrics and core numbers~\cite{av13, borgatti, wang, herland2013, tsugawa2015}.

Despite these studies, there is yet no definite answer to this key question -- {\em how can we identify whether a network would be stable under noise?}
Although studies such as ~\cite{wang} claim that the stability is affected by global properties, we see in the experiments reported here, that these correlations do not always hold. Instead, our observations lead us to the conclusion that it is the {\em local structure around the high centrality vertices, that significantly affect the stability}.

\noindent {\bf Noise model:} We consider an additive model, where a percentage of edges, selected from the complementary graph at random, are added to the existing network. We focus only on edge addition because missing edges can be predicted using link prediction algorithms. Therefore  understanding the effect of extraneous edges is the more critical problem.

\noindent {\bf Key observations and contributions:} We observe that two important factors affect the stability of the networks. First, {\em the stability of the rankings depend on the value of $k$}. The top-$k$ ranked vertices arrange themselves into clusters. Within a cluster, the centrality values are very similar. However, the difference in values between the last vertex of a cluster and the first vertex of the next cluster is large. This phenomena makes the distribution of the high centrality values look like a step function. If the value of $k$ falls at the end of a cluster, the results are stable, otherwise they become unstable. Second, we observe that {\em the stability of a network depends on the density of the subgraph induced by high centrality vertices, i.e., a rich-club}. This is because the centrality metrics are more affected by changes to their immediate neighbors than to nodes at a larger distance. 

These observations highlight that network stability is dependent on  localized subgraphs induced by the top-$k$ high centrality vertices, and not the global topology.  Our findings allow users to identify stable networks with respect to a centrality metric without applying the noise model on the network. {\bf Our main contributions} are as follows; 

\begin{titemize}
\item  Demonstrating that stability is dependent on the value of $k$ and that stability over successive $k$ is non-monotonic.
\item Demonstrating that stability of networks is high if the subgraph induced by the high centrality vertices is dense.
\item Providing a template to identify highly stable networks.
\end{titemize}

\section {Experiments and Observations} \label{exp}
Here we describe our experimental setup and report the behavior of the results under the perturbations. We used the following centrality metrics as defined in ~\cite{newmanbook}.

\begin{figure}[h]
\centering
    \includegraphics[width=3.5 in, height=1.5 in]{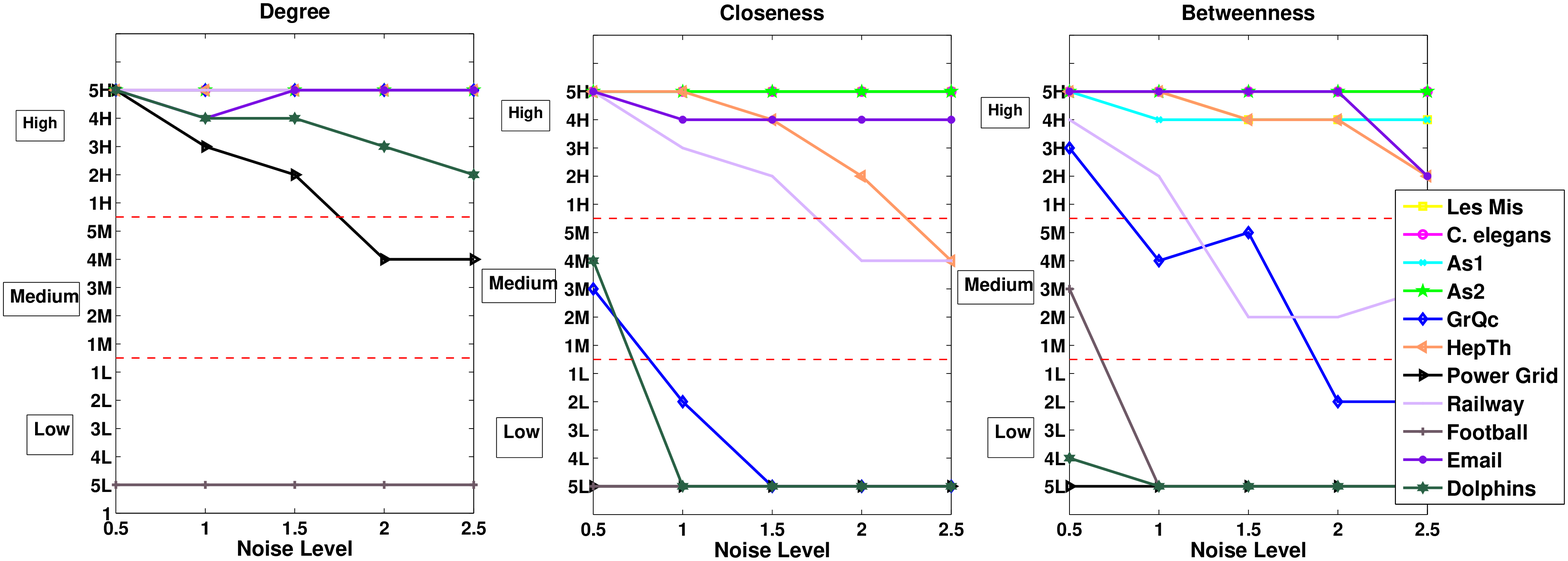}
    \caption{{\bf Changes in dominant stability of centrality metrics over different noise levels.} Left: Degree Centrality, Middle: Closeness Centrality , Right: Betweenness Centrality. X-axis: Noise Levels. Y-axis: Number of times the stability value fell in the high (H), medium (M) or low (L) range. {\bf The dominant stability decreases with increased noise level, but the rate of decrease depends on the centrality metric.}}
    \label{fig:summ}
\end{figure}
%\end{flushleft}
\begin{figure}
%\begin{tabular}{l}
\includegraphics[scale=.2]{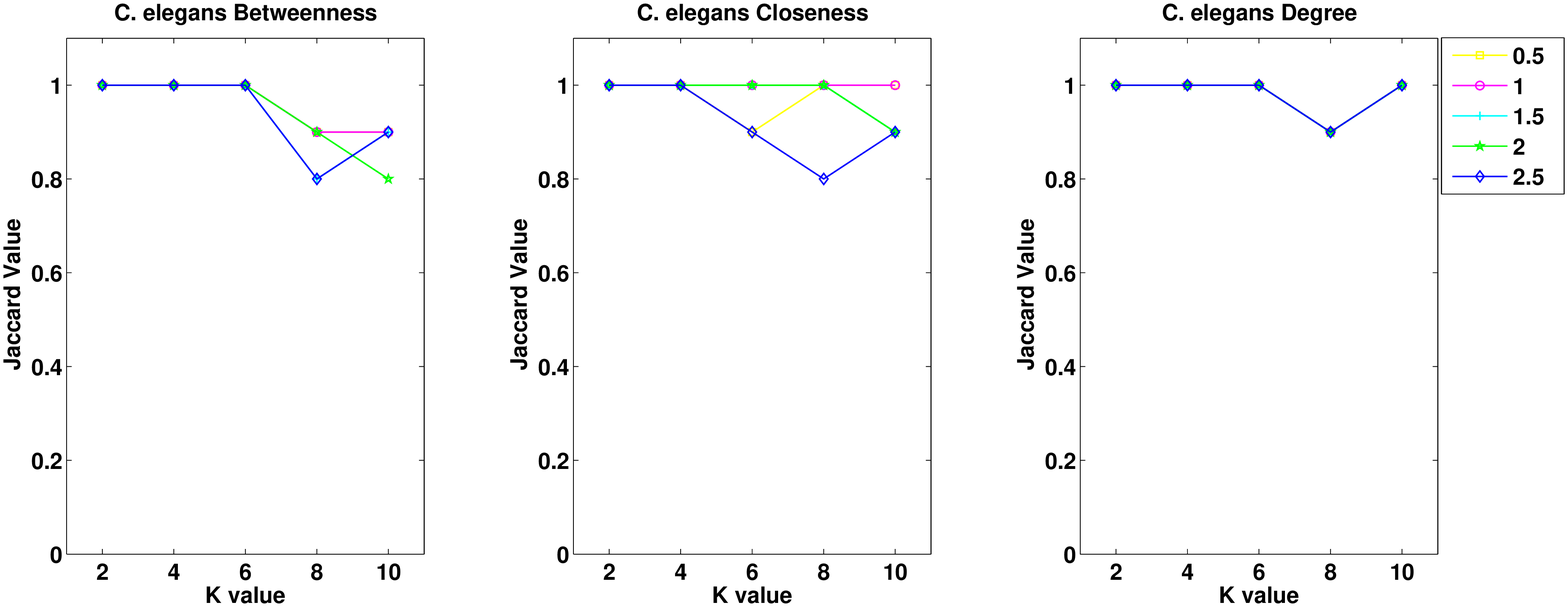} \\ \includegraphics[scale=.2]{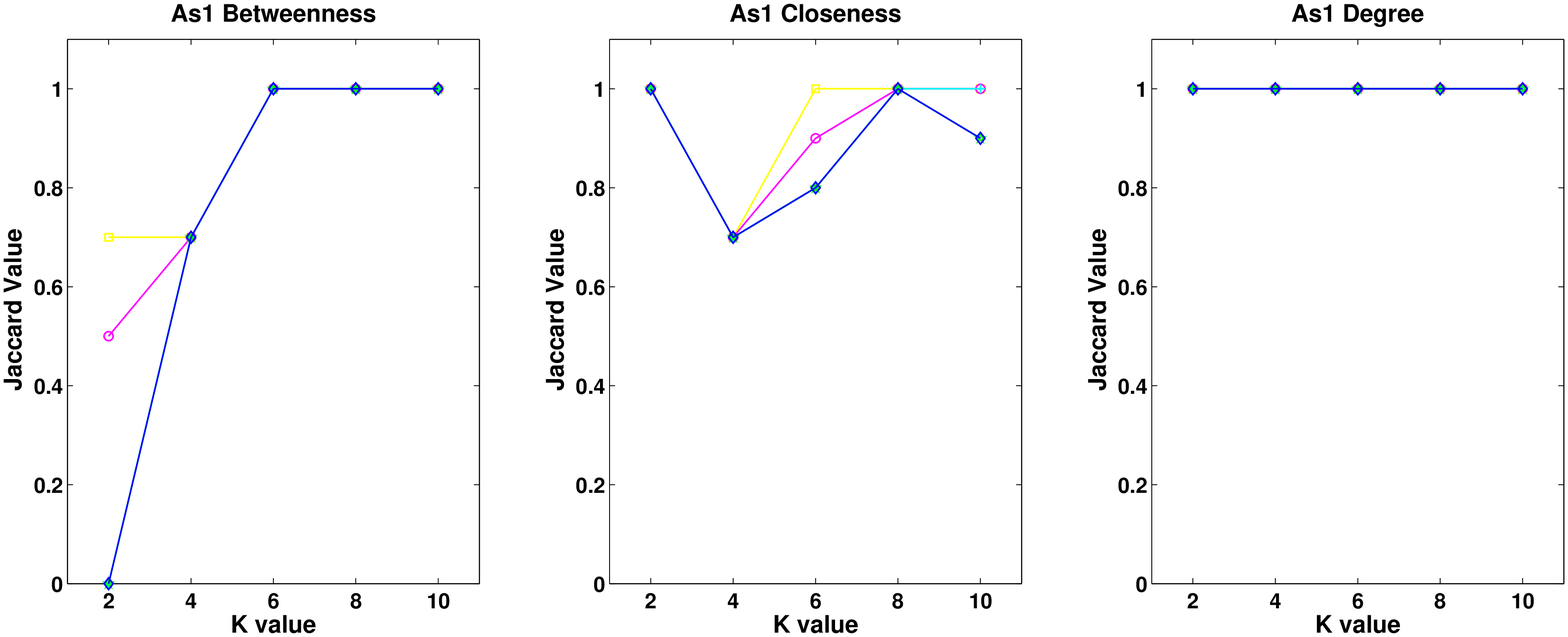}
\\
\includegraphics[scale=.2] {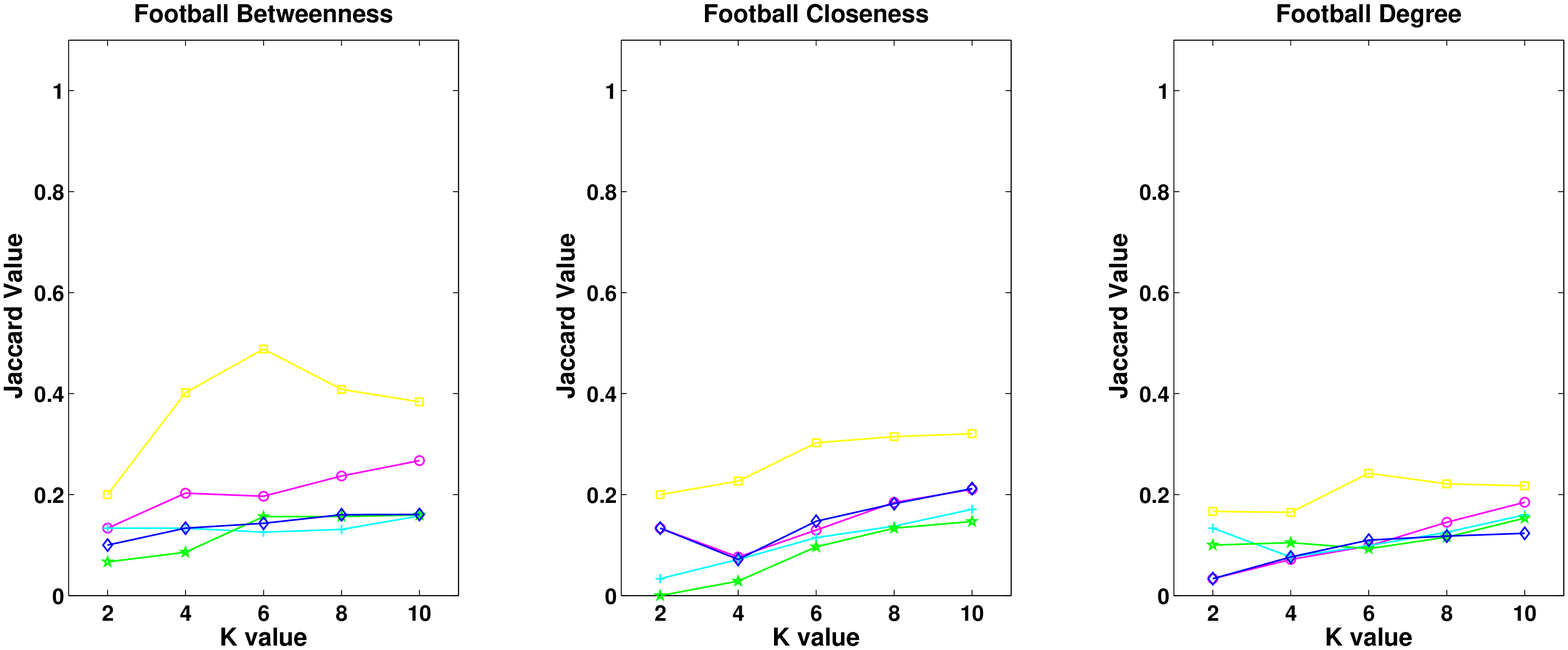} \\
\includegraphics[scale=.2]{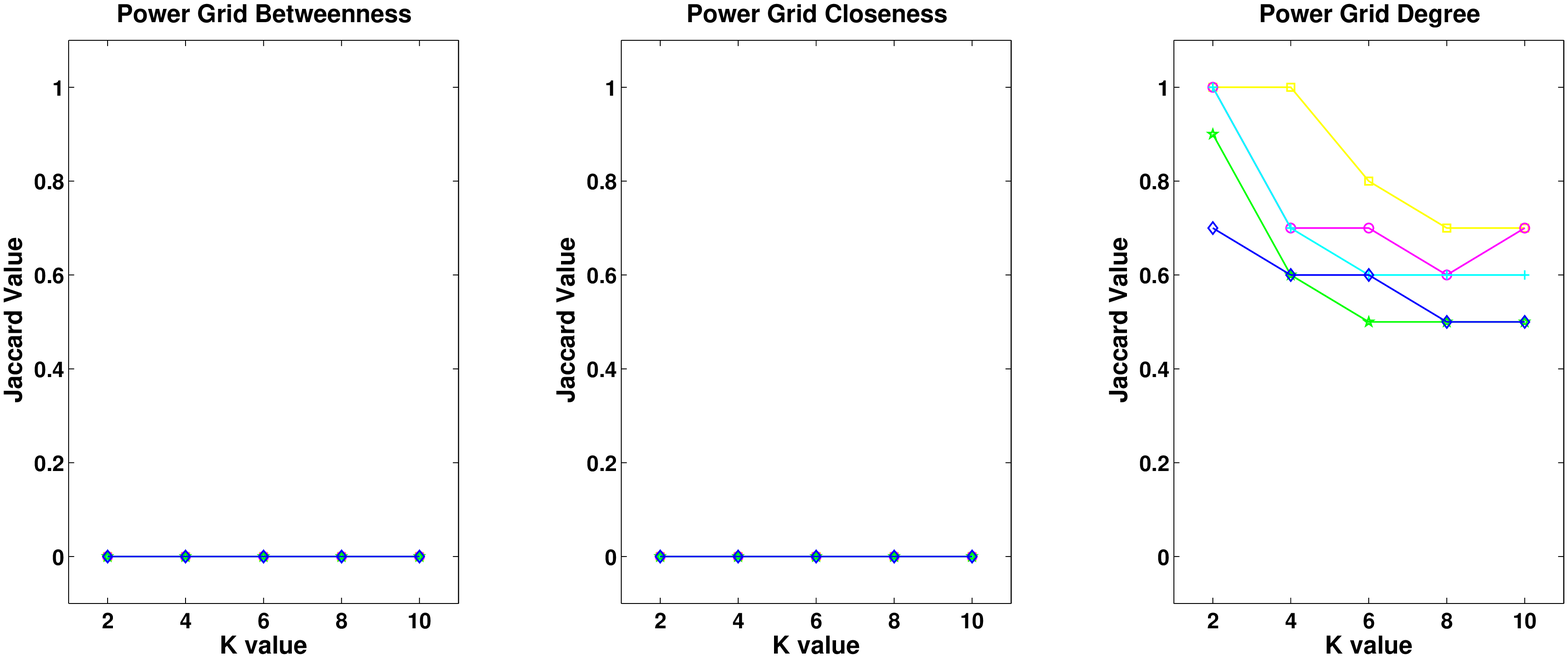} \\
\includegraphics[scale=.2] {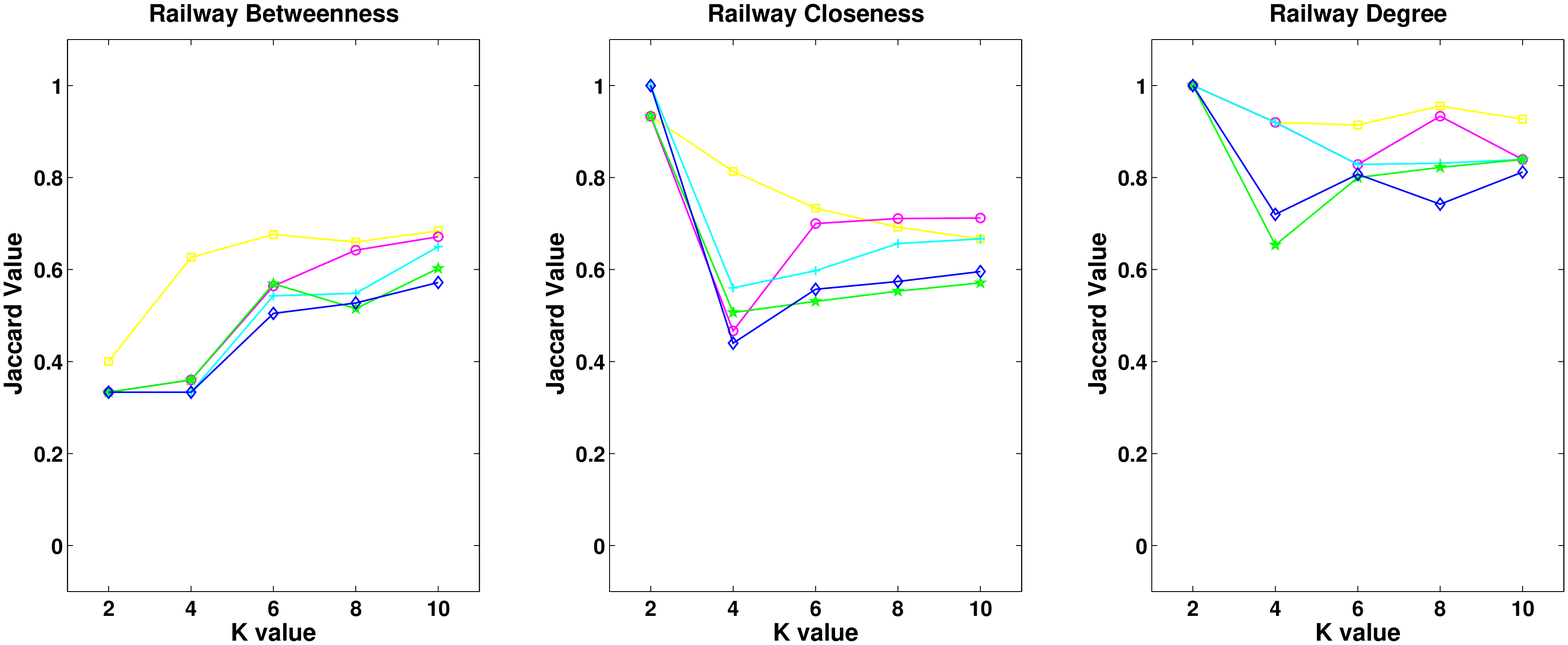} \\
\includegraphics[scale=.2]{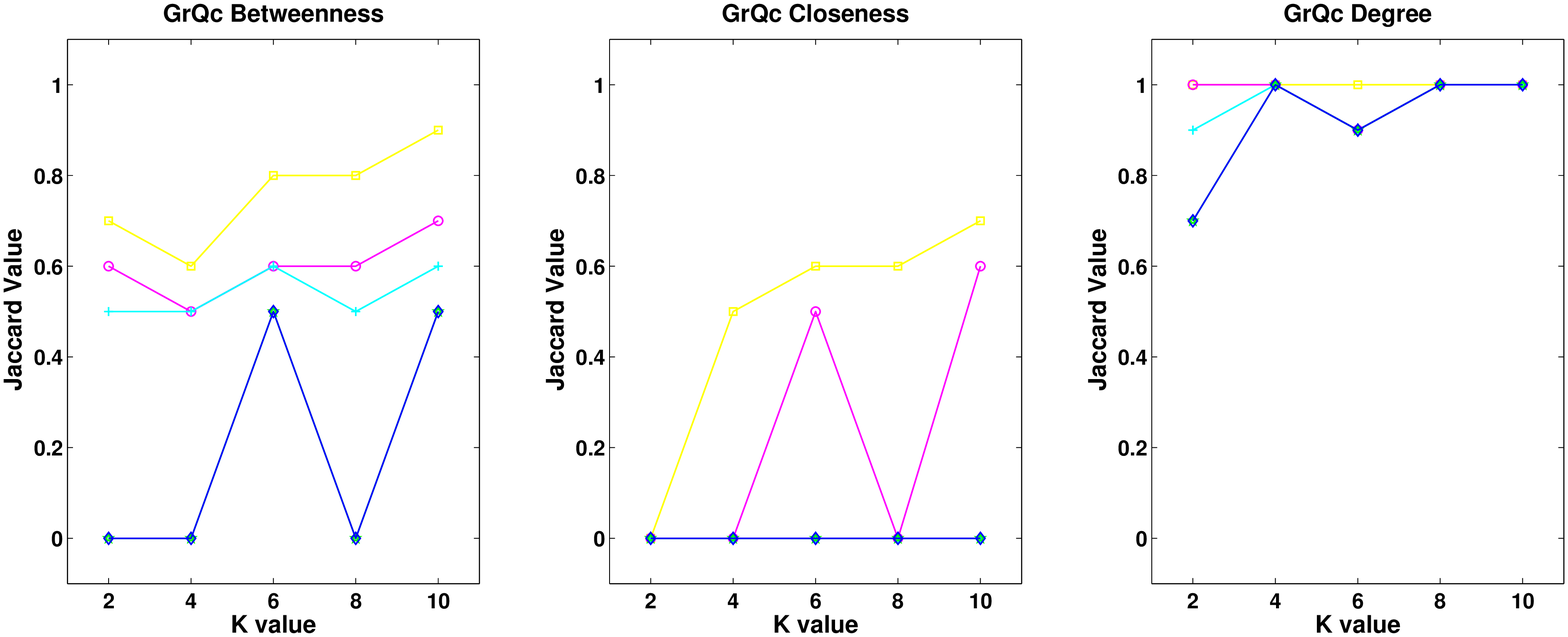}
%\end{tabular}
\caption{{\bf Change in centrality values for different $k$, over different noise levels.} Left: Betweenness, Middle: Closeness, Right: Degree. Y-axis: Jaccard Index. X-axis: The number of top vertices ($k$).  First two rows, networks with {\bf high stability}, Middle two rows, networks with {\bf low stability}, Last two rows, networks whose betwenness and closeness stability {\bf changes from high to low and back}. {\bf The choice of $k$ can significantly affect the JI value.}}
  \label{individual_netw}
\end{figure}

 {\em Degree centrality}, $D(v)$ of a vertex $v$ measures the number of its  neighbors. {\em Closeness centrality}, of a vertex $v$ is computed as  $CC(v) = \frac{1}{\sum\limits_{s\neq v \in V} dis(v,s)}$, where $dis(v,s)$ is the length of a shortest path between $v$ and $s$. {\em Betweenness centrality}  of a vertex $v$  is defined as 
$BC(v) = \sum\limits_{s\neq v\neq t \in V} \frac{\sigma_{st}(v)}{\sigma_{st}}$,where $\sigma_{st}$ is the total number of shortest paths between $s$ and $t$, and $\sigma_{st}(v)$ is the total number of shortest paths between $s$ and $t$ that pass through $v$.
We used the real-world networks listed in Table~\ref{table1} from ~\cite{snap, dim11, Ghosh}.

 \begin{table}[!htbp]
\centering 
 \caption{ {\bf  Test suite of real-world networks }}
  \begin{tabular}{|c|c|c|c|c|}
    \hline
    {\bf Name} & {\bf Nodes} & {\bf Edges} & {\bf CC} & $\alpha$\\ \hline
  %  { } & { } &{ }  &{\bf Coefficient} & {($\alpha$) } & {} \\ \hline      
%Karate   &  34  & 78 & 0.58 & 1.96 \\ \hline
%Chesapeake &   39 & 170 & 0.45 & 3.4 &  0.45\\ \hline
AS2 & 6474   & 13895 & 0.39 & 1.49  \\ \hline  
AS1 &   3570 & 7750 & 0.31 & 1.57 \\ \hline   
  C. elegans & 453 &  2025 & 0.65 & 1.65 \\ \hline  
                 
 Les Mis.  & 77  &  254 & 0.73 & 3.05 \\ \hline    
GrQc &   5242 &  14496 & 0.68 & 1.78 \\  \hline     
HepTh &   9877 &  25998 & 0.59& 1.66 \\   \hline  
Power Grid &   4941 & 6594 &0.10 & 1.45 \\   \hline    
Railway &   301 & 1224 & 0.74 & 6.68  \\   \hline 
Football &   115 & 613 & 0.40 & 1.57 \\   \hline 
Email &   1133 & 5451 & 0.25 & 2.75 \\   \hline 
Dolphins &   62 & 159 & 0.30 & 5.53 \\   \hline 
     \end{tabular}
  \label{table1}
    \end{table}

{\bf Noise model:} \label{noise}
  Our noise model is as follows. Of all the possible edges in a graph with $|V|$ vertices we pick an edge with probability  $\frac{\epsilon}{|V|}$ ; $\epsilon, 0\leq \epsilon \leq |V|$.   If the edge is not part of the network it is added to the network. 

If the degree of a vertex is $d$, then it has $n-1-d$ nodes that can get added due to perturbation.  Therefore the expected number of edges added to it will be $\frac{\epsilon}{|V|}(n-1-d)$. Thus  vertices with higher degree will have fewer edge additions.

{\bf Measuring stability:} For a given network and a given centrality metric we compute the stability as follows: We apply the noise model to the network with levels (i.e. $\epsilon$) of values $.5, 1, 1.5, 2, 2.5$.  For each network we compute the centrality values. We then compute the Jaccard Index (JI) to see how many of the top $k$ vertices in the original network are also among the top $k$ vertices in the perturbed network.  For two sets $A$ and $B$, $JI= \frac{A \cap B}{A \cup B}$. The highest value of JI is 1 (two sets have exactly the same elements) and the lowest value is 0 (sets have no elements in common). 

We conducted these experiments for each network over  10 perturbed networks per noise level. 
The JI presented in the results is the mean over the 10 networks. We  classified the stability into three groups ; {\it High Stability} (JI $ \ge  .7$); {\it Medium Stability} ($.4 \le $ JI $ < .7$) and {\it Low Stability} ($.4 > $JI $\ge 0$). 

{\bf Results:}
In Figure~\ref{fig:summ}, each line represents a network. The X-axis represents the noise levels ($\epsilon$).  For ease of visualization  we plot only for the even values of $k$ from 2 to 10.  Y-axis measures the {\em dominant stability}, i.e. the longest consecutively occurring stability range for that noise level. For example, 5H denotes that at that noise level, for all the five values of $k$ the stability was high. 3M denotes that for three consecutive values of $k$ the stability was in the middle range.

Figure~\ref{individual_netw} shows the changes for the individual networks, per value of $k$, not just the dominant stability. We have included two networks that were consistently in the high range (AS1 and C. elegans), two that were consistently in the low range (Power Grid and Football) and two that changed their stability values according to the centrality metric and noise level (GrQc and Railway). The results show that the stability value can change depending on the value of $k$.

{\bf Observations:}
The results show that even a small amount of noise (average edges added per vertex is 2.5) can significantly change the analysis results. However, the behavior of the three centrality metrics varies as follows:

\noindent{\em Degree:} The dominant stability decreases monotonically for {\em degree centrality}.  With the exception of Power Grid (has some middle level stability) and Football (all stabilities are low) all other networks show high stability. 

\noindent{\em Closeness:}  Several networks that show predominantly low dominant stability (Power Grid, Football, GrQc and Dolphins). Networks HepTh and Railway start as high stability, but their stability decreases with higher noise levels.

\noindent{\em Betweenness:} Here, Power Grid, Football and Dolphins have low dominant stability. GrQc goes from high, to medium to low. Railway also starts from high and ends at low. 

To summarize, our {\bf main observations} are as follows;
\begin{titemize}
\item The dominant stability decreases with increasing levels of noise. However, the individual stability changes non- monotonically with the values of $k$.
\item Among the centrality metrics, degree is most stable, closeness has a clearer separation between the high and low stability networks and in betweeness the separation is not as clear. The same network (e.g. GrQc at noise level 1.5) can have high (degree), low (closeness) and medium (betwenness) stability based on the centrality metrics.
\item The global topology of the network is not a deciding factor. In Table~\ref{table1}, the clustering coefficients  are very diverse and $\alpha$ is between 1.5 and 3. Neither of these parameters  seem to strongly correlate with the stability values.
\end {titemize}

\section{Factors Affecting Stability}\label{factors}
    \begin{figure}[htb]
\centering
\begin{tabular}{l}
\includegraphics[scale=.18]{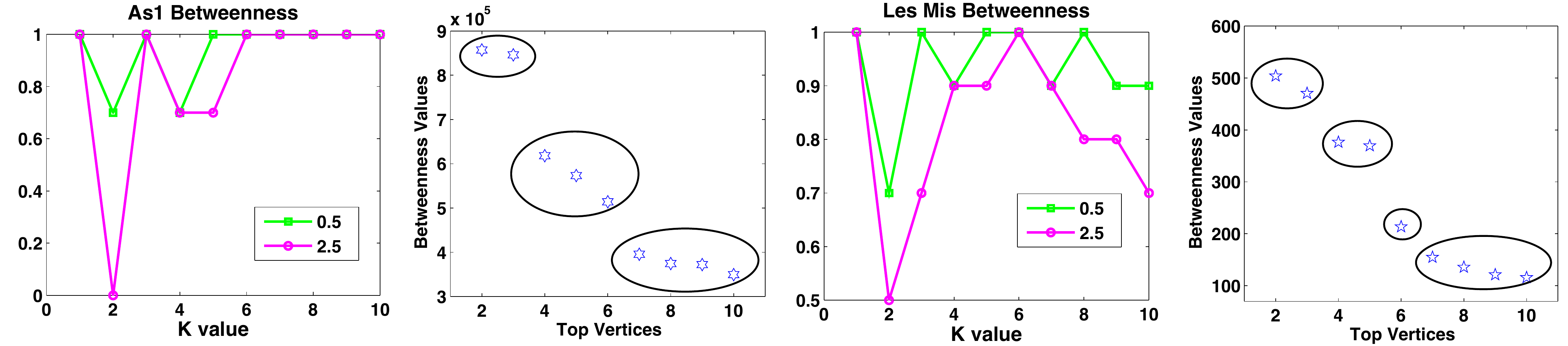}\\
\includegraphics[scale=.18]{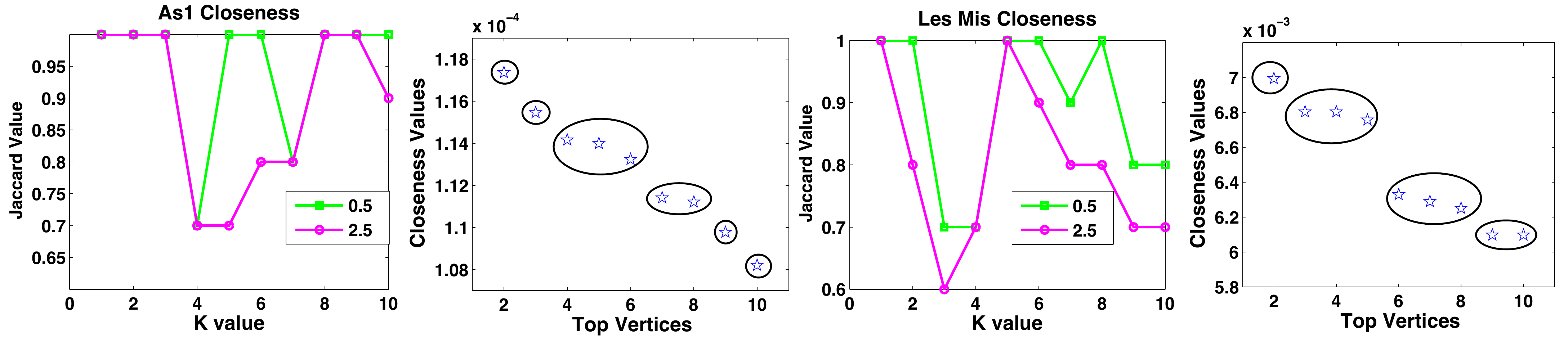}  
\end{tabular}
\caption{{\bf Stable clusters of centrality values} Top: Betweenness Centrality. Bottom: Closeness Centrality. {\bf Line Graphs:}  X-axis plots $k$, the number of top centrality vertices.  Y-axis plots the Jaccard Index. 
{\bf Scattered Plot:} X-axis plots vertex id for the top 10 centrality vertices. Y-axis plots the centrality values. The vertices  are clustered based on the relative difference of their centrality values. {\bf Stability increases when $k$ is in the beginning of a cluster and decreases when $k$ falls within a cluster}}

\label{fig:kvals}
\end{figure}
   
This section contains our {\em main contribution}, where we explain how properties of the network affect the stability. 

{\bf Stability of centrality metrics:}
\label{centrality_stability}
Consider two nodes $v_1$ and $v_2$, whose values for a  centrality metric, are  $X(v_1)$ and $X(v_2)$ respectively. In the original network, $X(v_1) >  X(v_2)$, thus $v_1$ has a higher rank than $v_2$~\footnote{We consider ranking from 1 (high) to $n$ (low). The vertex with highest centrality value is ranked 1}. After applying perturbation $p$, the centrality values  become $X_p(v_1)$ and $X_p(v_2)$. 

Our goal is to identify the lower bound on the difference between $X(v_1)$ and $X(v_2)$, such that after perturbation $X_p(v_1)$ will remain greater than $X_p(v_2)$. We consider the most optimal situation for $X_p(v_2)$ to become larger than $X_p(v_1)$. We assume that $X(v_1)$ has the maximum decrease  after perturbation and $X(v_2)$ has the maximum increase, given that  on average $\epsilon$  edges are added per vertex. Our computations for each centrality values are as follows;

\noindent{\bf Degree centrality:}  
The degree of a vertex will either increase or remain the same. Thus the maximum decrease of $v_1$ is zero. The value of $X_p(v_2) = X(v_2)+\epsilon$. Therefore, if $X(v_1)-X(v_2) >\epsilon$, then the ranking will not change.

For most networks the difference between the higher ranked vertices is larger than the maximum $\epsilon$ we set for our experiments, so the ranking of the vertices remain relatively stable.

\noindent{\bf Closeness centrality:} For simplicity, we consider $X(v)$ to be the inverse of closeness centrality, i.e. $X(v)= {\sum\limits_{s\neq v \in V} dis(v,s)}$

Since we are adding edges, this value will either increase  or remain the same.  The change in $X(v_2)$ will depend on where the edges are added.

Assume, due to perturbations, $v_2$ is added to a vertex $v_x$, which is at distance $d_x$ from $v_2$. Therefore, $v_x$, and other vertices whose shortest paths to $v_2$ passed through $v_x$ will have their distance to $v_2$ reduced by $d_x-1$. The maximum decrease is $X_p(v_2)= X(v_2)-\sum \limits_{t\in E_{add}}(d_t-1)R_t$, where $E_{add}$ is the set of nodes that are added to $v_2$, $d_t$ is distance of $t$ from $v_2$ in the original graph and $R_t$ is the number of vertices whose shortest path to $v_2$ passes through $t$. Thus the following has to hold: $X(v_2)-X(v_1) > \sum \limits_{t\in E_{add}}(d_t-1)R_t$ for the ordering between these two vertices to be stable.
$R_t$ will increase with $\epsilon$. The values of $d_t$ depends on the depth of the BFS tree originating from $v_2$.

{\bf Betweenness Centrality:} By adding edges the betweenness centrality of a vertex can increase if it gets connected to another high centrality vertex. It can also decrease, if new edges lead to alternate or smaller shortest paths.

Assume due to addition of edges to a vertex $v$, there are $R$ new pairs of vertices whose shortest paths pass through $v$. Also due to addition of edges in other parts of the network, there are $P$ pairs of vertices whose shortest paths used to pass through $v$ in the original network, but do not in the perturbed network. There are also $Q$ pairs of edges, whose length of shortest path does not change, but after perturbation there are new shortest paths between them.

We assume that $v_1$ sees only decrease in its BC value and $v_2$ sees only increase. Therefore $X_p(v_1)=X(v_1)-\sum\limits_{s_p\neq v\neq t_p \in P} \frac{\sigma_{s_pt_p}(v_1)}{\sigma_{s_pt_p}}
-\sum\limits_{s_q\neq v_1\neq t_q \in Q} \frac{q_x(\sigma_{s_qt_q}(v_1))}{(\sigma_{s_qt_q}+q_x) \sigma_{s_qt_q}}$, where $q_x$ is the number of new shortest paths for the vertex pair $s_q$ and $t_q$  and $X_p(v_2)=X(v_2)+\sum\limits_{s_r\neq v_2\neq t_r \in R} \frac{\sigma_{s_rt_r}(v_2)}{\sigma_{s_rt_r}}$.  
Therefore, the difference between $X(v_1)-X(v_2)$, must be larger than \\ $\sum\limits_{s_p\neq v\neq t_p \in P} \frac{\sigma_{s_pt_p}(v_1)}{\sigma_{s_pt_p}} +
\sum\limits_{s_q\neq v_1\neq t_q \in Q} \frac{q_x(\sigma_{s_qt_q}(v_1))}{(\sigma_{s_qt_q}+q_x) \sigma_{s_qt_q}}\\+\sum\limits_{s_r\neq v_2\neq t_r \in R} \frac{\sigma_{s_rt_r}(v_2)}{\sigma_{s_rt_r}}$

The number of elements in $R$ will increase as $\epsilon$ increases. The number of elements in $P$ and $Q$ depend on the length of the shortest paths. If the length of most of the shortest  paths through $v_1$ is already low,  then there is less chance that they will become even shorter or alternate paths will be found with addition of new edges. Based on these formulas we observe that the {\bf stability decreases with higher $\epsilon$.} For closeness and betweeness centrality, the increase (if any) also depends on network structure.

{\bf Stability based on centrality values:}  \label{cluster}
We observe that the relative differences of consecutive centrality values can indicate whether the ordering will be maintained.  
Figure~\ref{fig:kvals} plots the change in stability  as the noise levels remain constant, and the value of $k$ changes (line-graphs) and the values of the top-10 high centrality vertices (scattered plots).\footnote{Rank 1 node is not shown due its high value. By plotting it, the relative difference between the other vertices cannot be visualized well.}
The vertices can be grouped into clusters, where within the clusters the values are relatively close to each other, and across the clusters there is a large difference between the last vertex in the previous cluster and the first vertex in the next cluster. The stability increases when $k$ is at the beginning of a cluster ($k=4$ for AS1 Betweenness) and decreases when $k$ is within the cluster ($k=3$ for AS1 Betweenness).
 \begin{figure}[htb]
\centering
%\begin{tabular}{l l}
\includegraphics[scale=.15]{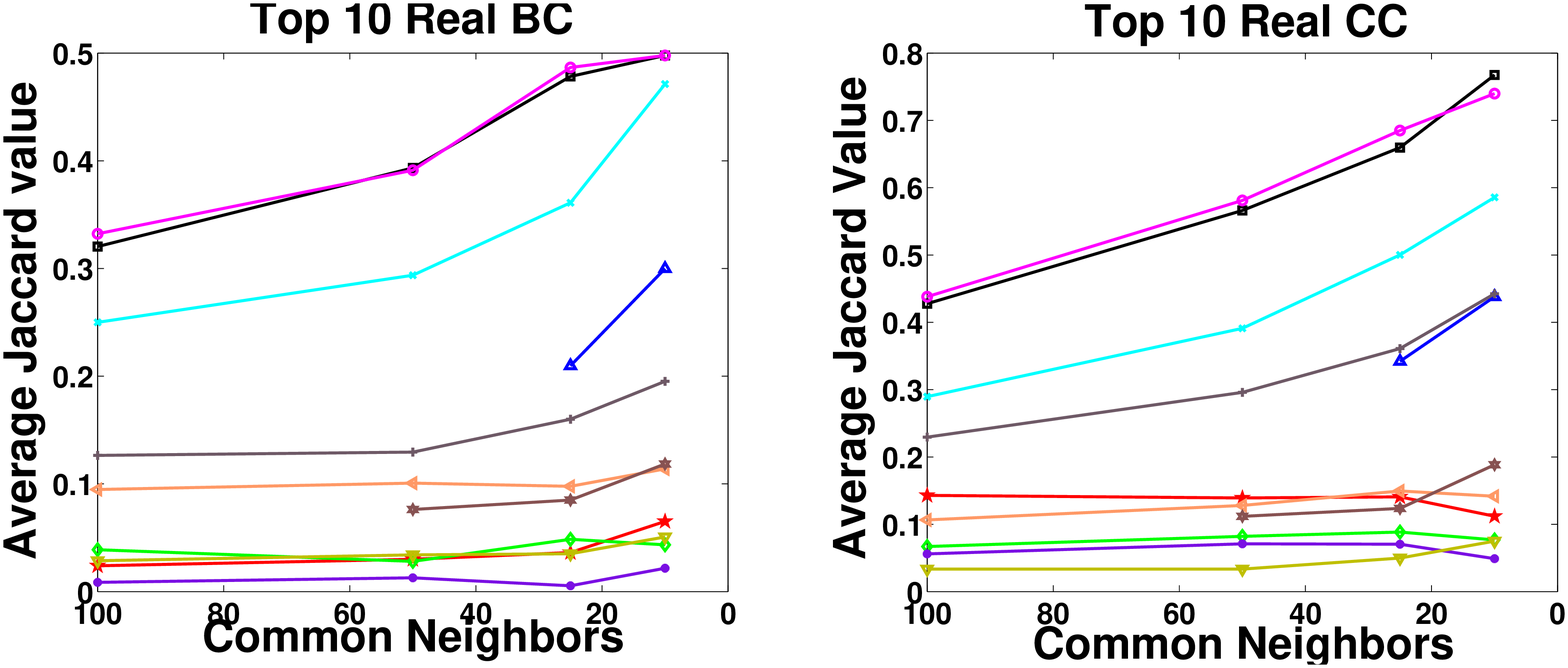} 
\includegraphics[scale=.15]{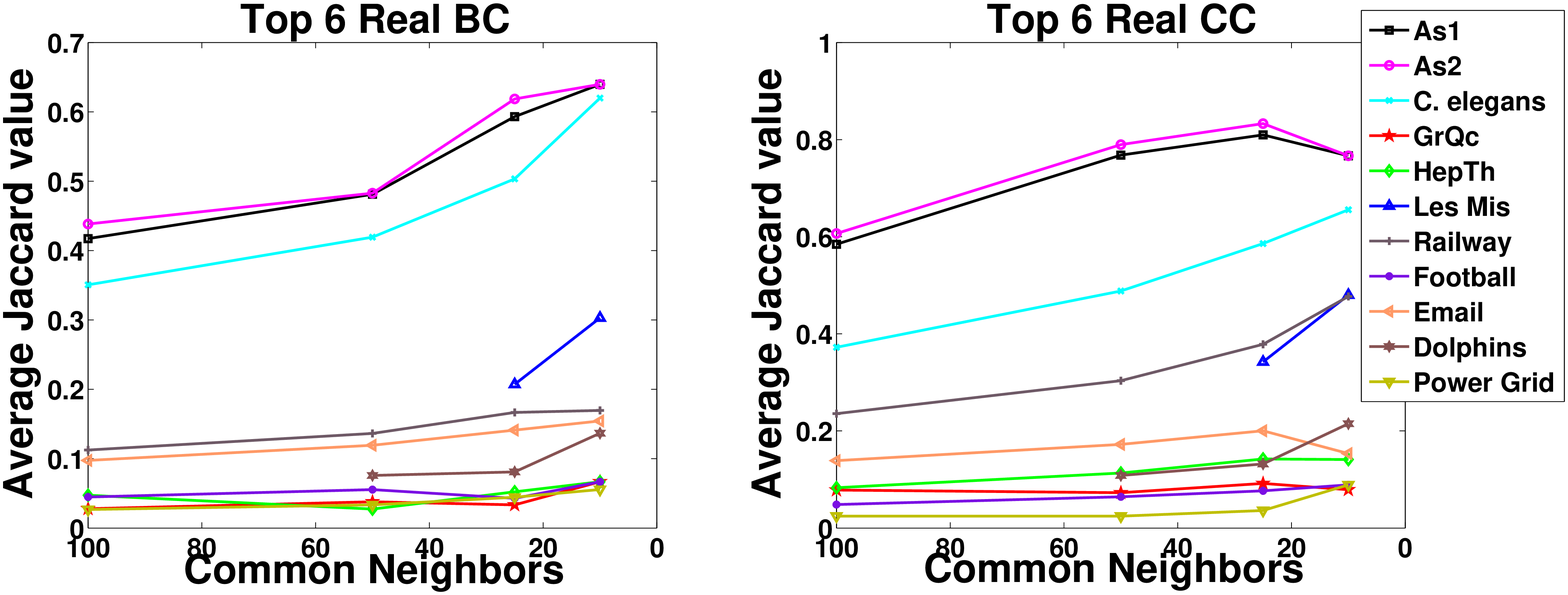}  
\includegraphics[scale=.15]{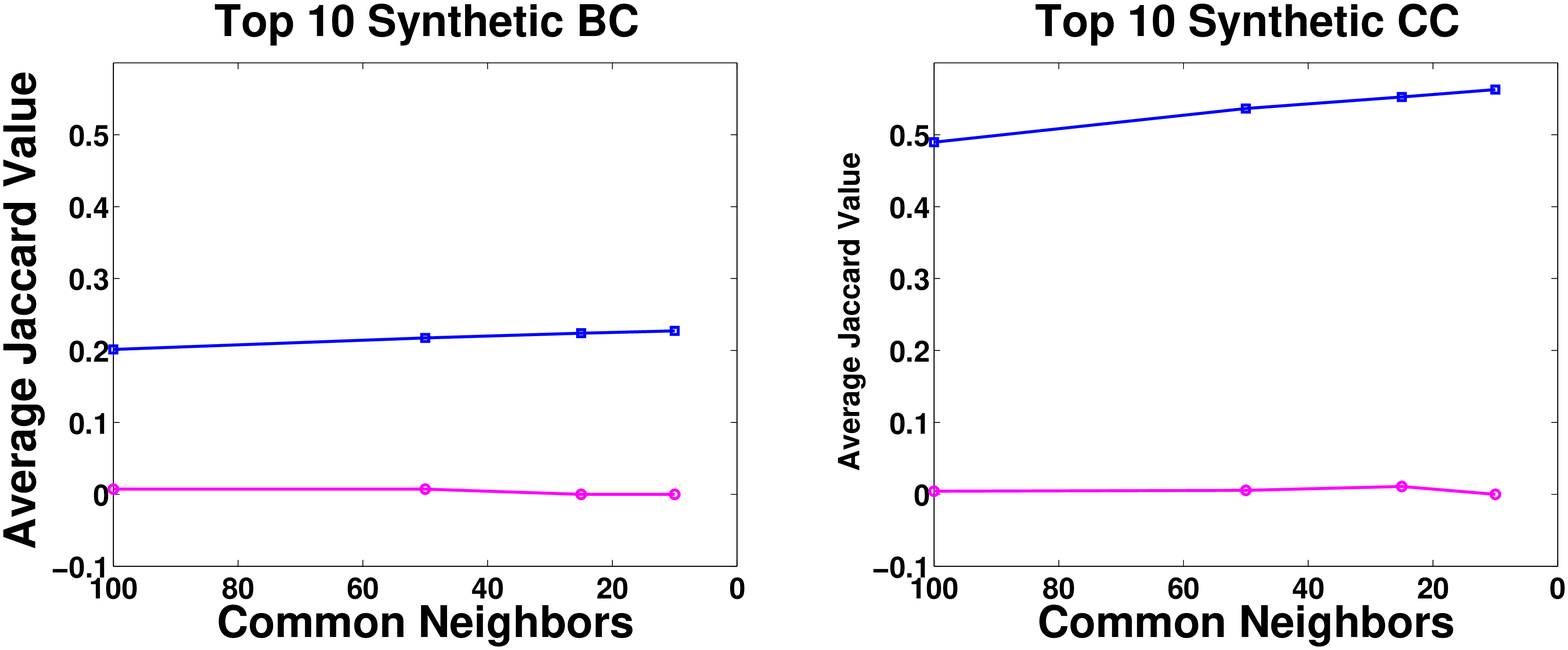}
\includegraphics[scale=.15]{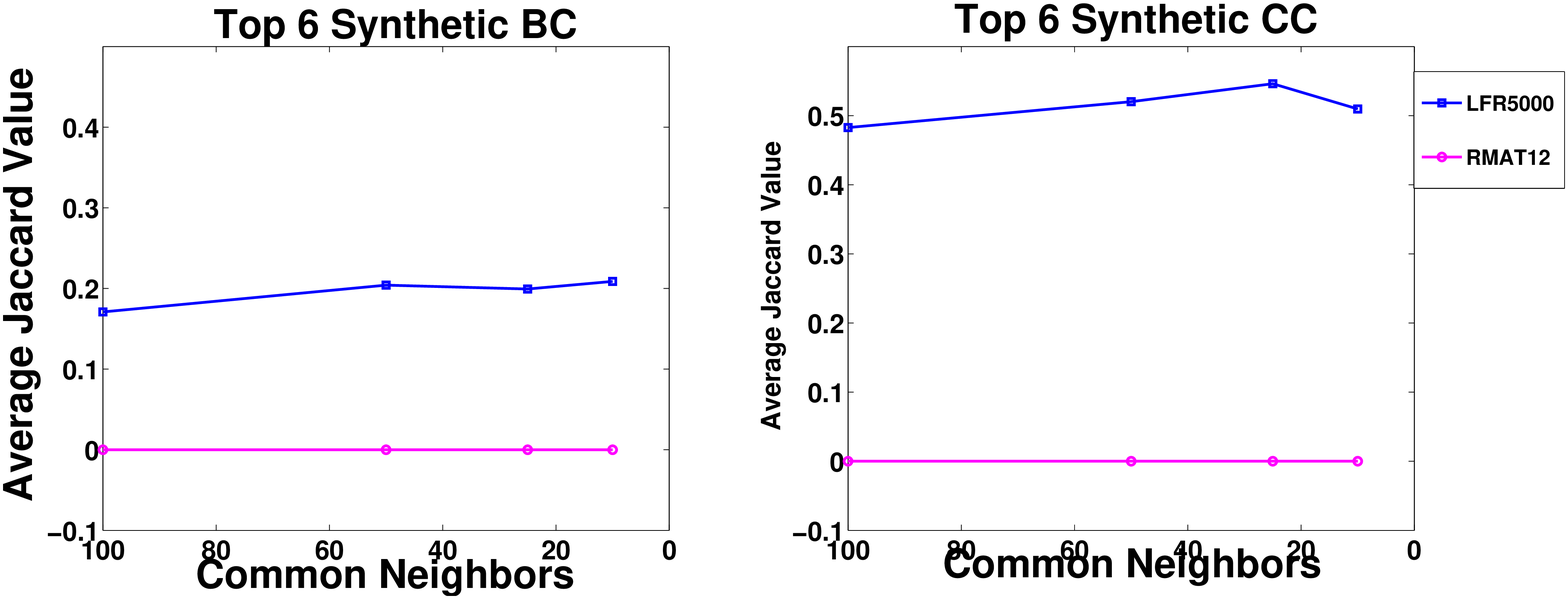}  
%\end{tabular}
 \caption{{\bf High ranked common neighbors of the top-$k$ ranked vertices}.
 X-axis is the number of high ranked neighbors. Y-axis is the average JI. For the high stability networks, the slope is increasing. This indicates that {\bf in stable networks, the most of the common neighbors of high ranked vertices are other high-ranked vertices.}}
 \label{commonN1}
\end{figure}

\noindent{\bf Stable clusters:} This phenomena occurs because it is  difficult to reverse the ranking between two vertices if they have a large difference in their values. However if the values are very close then slight perturbation can change the rankings. 

Therefore, we can use the relative difference between consecutively ranked vertices to group similarly valued vertices into {\bf stable clusters}. If the value of $k$ falls within the cluster, the Jaccard Index is likely to change. On the other hand, if $k$ is selected such that it falls at the beginning of the cluster, then  the ranking becomes more stable due to the large relative difference. This is borne out in Figure~\ref{fig:kvals}.

{\bf Identifying stable clusters:} To identify the stable groups we compare the difference between the centrality values of the consecutively ordered vertices. The breaks into clusters occurs between the two vertices that have the high relative difference. We continue dividing the vertices into clusters until the difference is lower than a certain threshold. Identifying these stable clusters allows us to have an improved understanding of how the network will behave under various levels of noise. {\bf Networks where the clusters are small in size and the clusters have high difference between them should have high stability.}

%\begin{flushleft}
\begin{figure} [htb]
\centering
\includegraphics[width = .8 in]{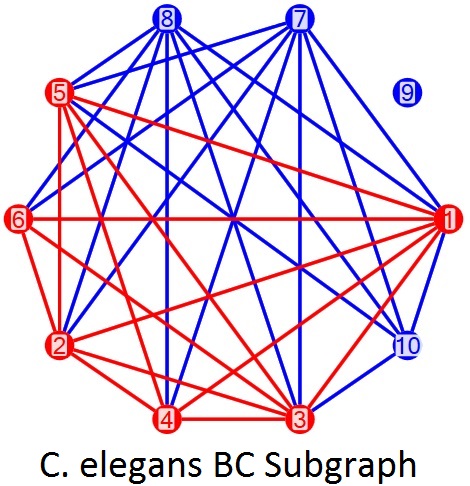}
\includegraphics[width = .8 in]{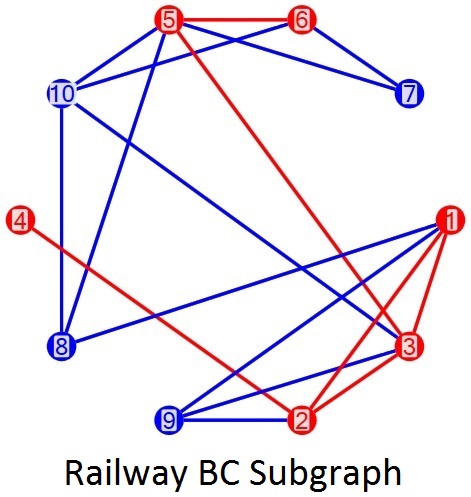}
\includegraphics[width = .8 in]{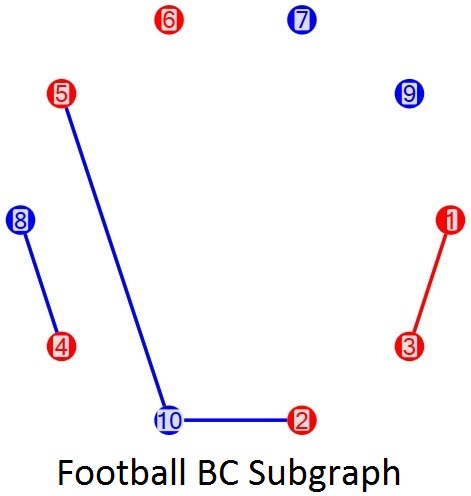}

\includegraphics[width = .8 in]{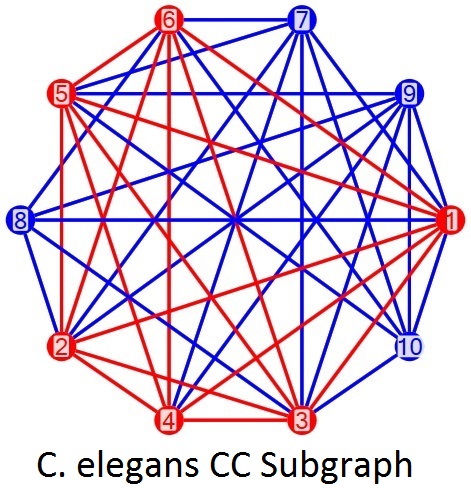}
\includegraphics[width = .8 in]{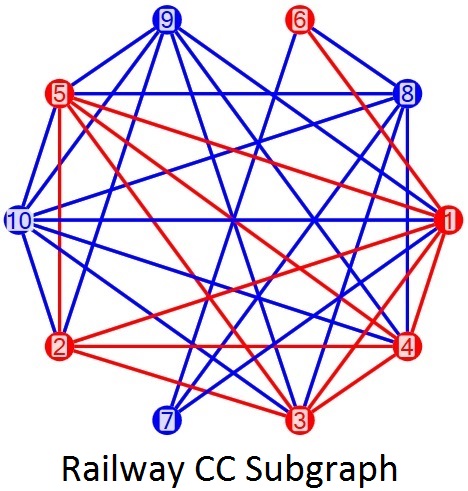}
\includegraphics[width = .8 in]{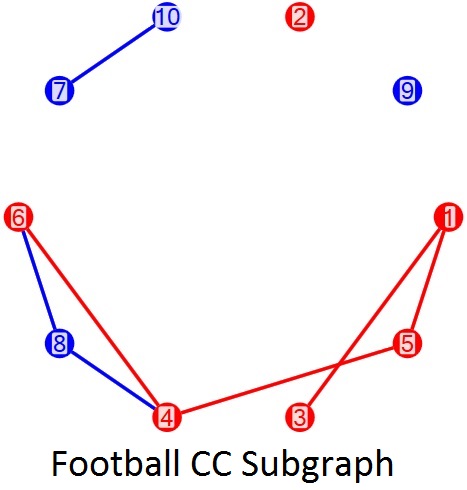}  
    \caption{ {\bf Subgraphs for high  betwenness (top) and high closeness (bottom) centrality vertices}. Left: C. elegans (high stability). Middle: Railway (medium stability). Right: Football (low stability). The red (blue) vertices and edges show the subgraph for the top 6 (10) vertices.  {\bf High stability networks have dense subgraphs and low stability networks have very sparse subgraphs.}}
    \label{subgraphs_all}
\end{figure}
%\end{flushleft}

 \begin{table*}[!t]%[!htbp]
\centering 
\caption{{\bf Comparing Stability and Local Connections of Networks} {\bf Network Stability} gives the mean of the stability over the noise levels at the specified $k$. {\bf Subgraph Density} gives the density of the subgraph induced by the top $k$ vertices. {\bf Common Top Neighbors} reports whether the corresponding line in Figure~\ref{commonN1} was in high, medium or low range. {\bf Networks with dense (sparse) clusters of high ranking nodes have high (low) stability.}}
 \begin{tabular}{|c||c|c||c|c||c|c|}
   \hline
 {\bf Network} & \multicolumn{2}{|c||} {\bf Network} & \multicolumn{2}{|c||}{\bf Subgraph}  & \multicolumn{2}{|c||} {\bf Common Top} \\  
   & \multicolumn{2}{|c||} {\bf Stability} & \multicolumn{2}{|c||}{\bf Density}  & \multicolumn{2}{|c||} {\bf Neighbors} \\ \hline 
 & {\bf Closeness} & {\bf Betweenness}  & {\bf Closeness} & {\bf Betweenness} & {\bf Closeness} & {\bf Betweenness} \\ \hline
 \multicolumn{7}{|c|} {\bf Top 10 High Ranked Vertices } \\ \hline
 \multicolumn{7}{|c|} { Dense Cluster and High Stability Networks } \\ \hline
AS20000101   &   High (.96)  & High (1)  &  .97&  .71 & High & High \\\hline
AS20000102    &  High (1)   & High (.78)  & .95 & .71 & High & High \\ \hline   
 C. elegans   & High (.94)   &  High (.76)  & .82 & .66 & High & High \\ \hline  
 
   Les Mis    &   High (.8)   &  High (.76) &  .66 &  .46 &
   Medium & Medium \\ \hline 
   
 \multicolumn{7}{|c|} { Sparse Cluster and Low Stability Networks } \\ \hline
GrQc  &   Low (.26)  &  Medium (.64)  & .26 & .11 & Low & Low \\   \hline
 Dolphin  &  Low (.1)  &  Low (.1) & .36 & .24 & Low &  Low\\   \hline
Football  &  Low (0)    &  Low (0) & .16 & .09 & Low & Low \\   \hline
Power Grid   &    Low (0) &  Low (0) & .24 & .15 & Low& Low \\   \hline 
   \multicolumn{7}{|c|} { Outlier Networks } \\ \hline
   Email   &  High (.98)   & High (.96)  & .31 & .24 & Low &  Low\\ \hline
  Railway    &   Medium (.68)  &  Medium (.68) & .67 & .38 & Medium &  Low \\ \hline
  
HepTh   &  Medium (.68)    &  High (.72) & .17 & .13 & Low & Low \\   \hline 
\multicolumn{7}{|c|} { Synthetic Networks } \\ \hline
LFR5000   &   High (.78)  & High (1)  &  .77&  .44 & High & Medium\\ \hline
RMAT12    &  Medium (.58)   & Medium (.48)  & .04 & .04 & Low & Low \\ \hline

 \multicolumn{7}{|c|} {\bf Top 6 High Ranked Vertices } \\ \hline
   
      \multicolumn{7}{|c|} { Dense Cluster and High Stability Networks } \\ \hline
AS20000101   &   High (.86)  & High (1)  &  1&  .93 & High & High\\ \hline
AS20000102    &  High (.8)   & High (.96) & 1 & .93 & High & High \\ \hline   
 C. elegans   & High (.96)    &  High (1) & 1 & .87 & High & High \\ \hline  
 
   Les Mis    &   High (.90)  &  High(1) &  .87 &  .60 &
   Medium & Medium \\ \hline 
   \multicolumn{7}{|c|} { Sparse Cluster and Low Stability Networks } \\ \hline
   GrQc  &   Low (.22)  &  Medium (.60) & .20 & .13 & Low & Low \\   \hline
Dolphin  &  Low (.12)   &  Low (0) & .47 & .40 & Low &  Low\\   \hline
Football  &  Low (0)   &  Low (.10)& .27 & .07 & Low & Low \\   \hline 
Power Grid   &    Low (0) &  Low (0) & .27 & .13 & Low& Low \\   \hline  
   \multicolumn{7}{|c|} { Outlier Networks } \\ \hline
   Email   &  High (.90)   & High (.76) & .33 & .27 & Low &  Low\\ \hline
  Railway    &   Medium (.68)  &  Medium (.62)& .73 & .40 & Medium &  Low \\ \hline
   
HepTh   &  Medium (.66)   &  High (1) & .27 & .13 & Low & Low \\   \hline   
\multicolumn{7}{|c|} { Synthetic Networks } \\ \hline
LFR5000   &   High (.72)  & High (.84)  &  .73 &  .4 & High & Medium\\ \hline
RMAT12    &  Medium (.64)   & Medium (.6)  & .07 & .07 & Low & Low \\ \hline  
    \end{tabular}
 \label{subgraphtable}
   \end{table*}

{\bf Stability based on network structure:}\label{net_vary}
We now see how the network structure affects its stability.  The slope of the degree distributions ($\alpha$) for most  of the networks in our test suite are from 1.5 to 3, their average local clustering coefficient is very varied, and neither of these  correlate to the stability of the networks. 
Therefore, as seen from the earlier equations that the {\bf stability seems to be dependent on the local structure of the high centrality vertices.}

\noindent{\bf High ranked common neighbors:}  We  investigated whether the common neighbors of the top $k$  nodes also have high rank. For each pair of nodes within the top $k$  ($k$=10 and $k$=6) set (for a given centrality metric) we calculated the Jaccard Index between their connections to the top 100,  50,  25, and  10 high ranking nodes, and  computed the average JI for each set of neighbors (top 100, top 50, etc.). 

As the range of high ranked neighbors decreases (from 100 down to 10), the average JI value increases (Figure~\ref{commonN1}).  This indicates that the {\bf top-$k$ high rank nodes have more common neighbors among the high-ranked nodes}.  The curves divide into three regions.  The top networks are the ones with high stability (e.g., C. elegans), the networks in the middle have not so high stability (e.g., Email) and the ones at the bottom show low stability (e.g., Football). 

\noindent{\bf Subgraph induced by high ranked vertices:}  For each metric, we identified the top $k$ high ranked vertices and then computed the density of the induced subgraphs from the vertices in this set.  {\bf Networks that achieve more instances of high stability have more dense subgraphs} (Figure~\ref{subgraphs_all}).
 
\noindent{\bf Summary:} Table~\ref{subgraphtable} summarizes the results for $k=10$ and $k=6$. The density of a subgraph is  the ratio of the total number of edges in the subgraph by the total possible edges. 

If the networks have high stability for the top-$k$ vertices, then the subgraph induced by those vertices is also dense ($\ge$ .60). Conversely, if the network has low stability, then the corresponding subgraphs are sparse ($\le$ .40, with Dolphin being the exception). This pattern is also observed when comparing their common neighbors. For high (low) stability networks, the corresponding line in Figure~\ref{commonN1} is in the high (low) range. The results are similar for $k=10$ and $k=6$.

The exceptions are listed under Outliers. For example, Email and HepTh have high stability but low subgraph density. In these cases, a smaller subgroup of the high centrality vertices form a dense cluster, and the remaining high centrality vertices connect to that cluster. Another case is Railway with medium, tending to high stability for BC.  Here, the subgraph for betwenness consists of two smaller clusters connected to each other (see Figure~\ref{subgraphs_all}). Similar characteristics appear for Les Mis and GrQc (high BC, low density) 

Table~\ref{subgraphtable}, also shows the results of two synthetic networks,  RMAT12 (random network created using RMAT generator) and LFR5000 (scale free network created using LFR generator with $\mu$=.1). The subgraph density of RMAT12 is constant for both centralities.  LFR5000 however, shows strong subgraphs for closeness and  a strong cluster over a subset of vertices for betweenness centrality. Therefore {\bf compared to random graphs, scale-free networks with strong communities are more stable.} 

{\bf Template to detect high stability networks:} Based on our observations, we propose a template to identify stable networks as follows;
%\begin{enumerate}
 1. Identify the top-$k$ centrality nodes and  their values
2. {\em Stability Condition 1:} Identify the lower bound between the differences of the centrality values that will maintain the ordering. If the difference in the high centrality nodes is greater than the lower bound, then the network is stable for that range of $k$.

3. {\em Stability Condition 2:} Find stable clusters based on the values of high centrality nodes. If $k$ falls at the beginning of the cluster, then the network is stable for that range of $k$.

4. {\em Stability Condition 3:} Find the subgraph induced by the top-$k$ nodes. If the subgraphs are dense and the number of common high ranked neighbors is high, then the network is stable for that range of $k$.
%\end{enumerate}

If all these conditions are satisfied, the network should be highly stable. Conversely, if none of these conditions are satisfied the network should have low stability. Note that our method does not require the user to actually perturb the network to estimate its stability.

\section{Discussion} \label{discussion}
Our experiments demonstrate two extremely important findings which have so far never been observed. The first is that networks where the high centrality vertices are very well-connected, i.e., they form a ``rich-club'', are more stable. The second is that the stability of the rankings of nodes depends on the number of top ranked nodes ($k$) being investigated. The top nodes seem to arrange themselves into groups; if the value of $k$ is such that it does not split a group then the results are stable, otherwise they are unstable. 

  Based on  these conditions of stability, users can evaluate the stability of their networks, without applying the noise model. They can also use these conditions to improve the stability of their data collection methods.

In future, we plan to extend this study to other forms of noise models and other varieties of network properties. We also plan to develop methods to determine the thresholds automatically. A final direction would be to analyze the performance of the stability detection algorithm for  other networks and  other application areas.

\section{Acknowledgements} Sandjukta Bhowmick and Vladimir Ufimtsev  thank NSF Award Number 1533881 for funding this project.

\bibliographystyle{abbrv}
\bibliography{sbhowmick1,Ufimtsev}
% You must have a proper ".bib" file
%  and remember to run:
% latex bibtex latex latex
% to resolve all references

\end{document}